# Quantum-like representation of neuronal networks' activity: modeling "mental entanglement"

Andrei Khrennikov[1,*], Makiko Yamada[2,3,4]

[1] Center for Mathematical Modeling in Physics and Cognitive Sciences
Linnaeus University, Växjö, SE-351 95, Sweden
[2] Institute for Quantum Life Science
National Institutes for Quantum Science and Technology, Chiba, 263-8555, Japan
[3] Institute for Quantum Medical Science
National Institutes for Quantum Science and Technology, Chiba, 263-8555, Japan
[4] Graduate School of Science and Engineering
Chiba University, Chiba, 263-8522, Japan
*Corresponding author email: Andrei.Khrennikov@lnu.se

**Abstract**: *Quantum-like modeling* (QLM) – quantum theory applications outside of physics – are intensively developed with applications in biology, cognition, psychology, and decision-making. For cognition, QLM should be distinguished from quantum reductionist models in the spirit of Hameroff and Penrose and well as Umezawa and Vitiello. QLM is not concerned with just quantum physical processes in the brain but also QL information processing by macroscopic neuronal structures. Although QLM of cognition and decision-making has seen some success, it suffers from a knowledge gap that exists between oscillatory neuronal network functioning in the brain and QL behavioral patterns. Recently, steps toward closing this gap have been taken using the generalized probability theory and prequantum classical statistical field theory (PCSFT) – a random field model beyond the complex Hilbert space formalism. PCSFT is used to move from the classical ``*oscillatory cognition*'' of the neuronal networks to QLM for decision-making. In this study, we addressed the most difficult problem within this construction: QLM for entanglement generation by classical networks, i.e., *"mental entanglement."* We started with the observational approach to entanglement based on operator algebras describing "local observables" and bringing into being the tensor product structure in the space of QL states. Moreover, we applied the standard states entanglement approach: entanglement generation by spatially separated networks in the brain. Finally, we discussed possible future experiments on "mental entanglement" detection using the EEG/MEG technique.

**keywords:** Quantum-like modeling; neuronal networks; mental entanglement; decision making; EEG/MEG technique

## 1. Introduction

Intensive development of quantum information theory has transformed the perspective of quantum studies toward an information-based approach to physics. In particular, numerous information-theoretic interpretations of quantum mechanics have been proposed [1]. These interpretations exert both foundational and technological influence. This informatization of quantum physics has also stimulated applications of the methodology and formalism of quantum theory beyond physics, extending to biology, cognition, psychology, decision-making, economics, finance, and the social and political sciences (see, e.g., monographs [2]-[8] and reviews [9,10])—a direction commonly termed *quantum-like modeling* (QLM). In this paper, we focus on QLM in the domains of cognition and decision-making.

Here, QLM must be clearly distinguished from quantum reductionist models advanced by Hameroff and Penrose [11,12], Vitiello [13,14], and Igamberdiev [15,16], who associated cognition and consciousness with quantum physical processes in the brain. Hameroff and Penrose emphasized microtubules, Vitiello referred to quantum field theory and long-range correlations in the brain, and Igamberdiev linked cognition to quantum processes in cells. In contrast, QLM of cognition does not concern quantum physical processes in the brain but rather quantum-like information processing by macroscopic neuronal networks.

QLM of cognition and decision-making has been successfully developed; it mathematically describes non-classical features of cognitive phenomena, such as "interference and entanglement of minds." It resolves numerous paradoxes of decision theory and models basic cognitive effects, including conjunction, disjunction, order, response replicability, and Zeno effects (see the mentioned works and articles [17]-[23]). QLM has highlighted the contextuality of cognitive phenomena by applying advanced quantum contextuality theory and, in particular, the machinery of Bell inequalities [24]-[29] (Section 11]). QLM has introduced a novel perspective on rationality versus irrationality and violations of the Savage Sure Thing Principle [4], framed within probabilistic and logical approaches, as violations of Bayesian updating [21,22], the formula of total probability [2,30,3,,6,31], and classical Boolean logic, while incorporating quantum logic [32,33,34]. QLM has successfully described statistical data on bistable perception of ambiguous figures 17,35,36,37, 2], has been applied to biological evolution (including genetic and epigenetic mechanisms) [40,6,41], and, more recently, to aesthetic experiences during book reading [42,43].

Nevertheless, cognitive QLM faces a gap between theoretical descriptions of classical oscillatory neuronal networks in the brain (e.g., the phase space description of harmonic oscillators) and the quantum-like representation of mental states and observables (e.g., density and Hermitian operators). Thus, it remains a phenomenological framework.

Recently, progress toward bridging this gap has been achieved [44] within the framework of *prequantum classical statistical field theory* (PCSFT)—a random field model generating the complex Hilbert space formalism for states and observables [45]-[48]. PCSFT has been employed to connect classical "oscillatory cognition" of neuronal networks with QLM for decision-making.[i1]

---

[1] See [49]-[54] for other models aimed at coupling neuronal and QL information processing in the brain. We especially highlight the article [55] in that we employ *generalized probability* (operational measurement) theory. This is a more general formalism than quantum theory in a complex Hilbert space. But all authors exploring QLM work within the

In this paper, we proceed to the most difficult problem within such construction: creation of QLM for the *generation of entanglement by classical networks*–the problem of *"mental entanglement."*

We now outline the content of the paper. Section 2 presents the basic construction for transitioning from oscillatory dynamics of neuronal networks in the brain to the quantum-like (QL) representation. Section 3 links classical and quantum realizations of observables on neuronal networks. Section 4 addresses a specific approach to the notion of entanglement, treating it as the entanglement of observables. In Section 5 this approach is applied to entanglement of observables on neuronal circuits. In Section 6 we examine the standard approach to entanglement as state entanglement and its application to modeling mental entanglement. In Section 7 we discuss the role of ephaptic coupling in generation of correlations between neuronal circuits in the brain. Section 8 concerns possible experimental verification of mental entanglement. This section deserves the special attention of experts in experimental cognitive science and neuroscience. Here, we discuss concrete experimental tests of mental entanglement based on classical electroencephalogram (EEG)/Magnetoencephalography (MEG) measurement techniques (Section 8.1), including comparative analysis with classical EEG-based approaches to functional connectivity in neuroscience (Section 8.3). Section 9 describes the main quantitative measures of entanglement that can be used in experimental tests. In Section 8.4, we analyze the possibility of creating and detecting entanglement in in vitro neuronal networks. Conceptual differences between classical and QL frameworks are discussed in Section 10. Section 11 provides a general discussion on the proposed model. Appendix A concerns the mathematical model for coupling classical oscillatory dynamics, represented by Hamiltonian equations, with quantum dynamics described by the Schrödinger equation. In Appendix B, we examine the possibility of detecting mental entanglement experimentally with Bell inequalities.

In physics, entanglement is one of the most intriguing and complex quantum phenomena. Typically, entanglement is treated as *state entanglement*. In the mathematical formalism of quantum theory, a pure state $|\psi\rangle$ of a compound system $S = (S_1, S_2)$ is given by a normalized vector belonging to the tensor product of two complex Hilbert spaces, $\mathcal{H} = \mathcal{H}_1 \otimes \mathcal{H}_2$. A state $|\psi\rangle \in \mathcal{H}$ is called entangled if it cannot be factorized, that is, it does not have the form $|\psi\rangle = |\psi\rangle_1 \otimes |\psi\rangle_2$, where $|\psi\rangle_i \in \mathcal{H}_i$, i=1,2.

Another less familiar approach to the notion of entanglement is *observation entanglement* [57-60] based on considering "local algebras" of cross-commuting observables $A_1$ and $A_2$. In this view, a tensor product structure on the state space is not preassigned but is generated by these operator algebras. The same state space $\mathcal{H}$ can be endowed with multiple tensor-product structures corresponding to different choices of operator algebras. In this paper, we employ both approaches to the notion of entanglement.

---

standard quantum formalism based on complex Hilbert space. So, it is useful to justify the use of this formalism from the neurophysiological viewpoint.

Since observation entanglement is less familiar and not widely applied, we complete the paper with a special section devoted to this notion, Section 4. We then proceed to the entanglement of observables on neuronal networks in Section 4.

State entanglement generated by a compound neuronal network $S = (S_1, S_2)$ is discussed in Section 6. Such entanglement and its generation by interacting neuronal networks is of neurophysiological interest. To identify its roots, we must look deeper into brain architecture and communication between neuronal circuits, including those not physically connected through axonal-dendritic pathways. Here, we emphasize the role of electromagnetic signaling, particularly ephaptic coupling between neuronal structures in the brain (Section 7). However, discussion of state entanglement in neuronal networks (Section 4) is primarily theoretical, as experimental detection remains far from reach. Observation entanglement (Section 6) is more promising for experimentation. The central challenge is identifying observables on neuronal networks capable of exhibiting such entanglement.

This *paper is conceptual* and aims to demonstrate the possibility of modeling generally nonlocal correlations in the brain that are mathematically described as quantum state entanglement. At present, the biophysical mechanism of generating such states is not completely clear (see, however, Section 7). This is the place to emphasize that "mental nonlocality," mathematically described as entanglement, is unrelated to the so-called spooky action at a distance often associated with "quantum nonlocality." Unlike quantum physics experiments on spatially separated systems, such as two photons 100 km apart, in cognitive science, we study a small physical system, the brain. Electromagnetic signals connect any two points in the brain almost immediately. Thus, *biological nonlocality expressed via entanglement is classical nonlocality generated by electromagnetic signaling between neuronal circuits*. Mental entanglement is the mathematical description of nonlocal correlations between observations performed on activated neuronal circuits. The crucial difference from classical correlations is that some observables in local algebras $A_i$ (associated with the neuronal networks $S_i, i = 1,2$) can be incompatible, not jointly measurable. In mathematical terms, such observables are described by non-commuting operators.

We note that in article [44], entanglement of neuronal networks was only mentioned in an unconventional framework that avoided tensor products, instead employing the classical Descartes product (see [45]-[48]). This approach may be interesting from the perspective of classical oscillatory cognition. However, by using it, we lose connection with the notion of entanglement as defined in quantum information theory.

We also speculate that by exploring the QL representation of mental states and observables, the brain may realize so-called *quantum-inspired algorithms* [61] and thereby achieve essential enhancement of computational power [62]. In such algorithms, the ability to generate entangled states is important.

We remark that QLM based on the QL representation of EEG signals is already applied in medical diagnostics of neurological disorders, including depression, epilepsy, and schizophrenia [63,64]. Such diagnostics work unexpectedly well, but in the absence of a theoretical justification. Mental entanglement can serve as the theoretical basis for EEG-based QL diagnostics, as it mathematically describes nonlocal information processing in the brain. The observed violation of the Clauser-Horne-Shimony-Holt (CHSH) inequality for EEG signals (transformed into dendrograms with clustering algorithms) can also be connected to mental entanglement.

## 2. QL states of neuronal networks

Let $S$ be a neuronal network with oscillatory node-circuits numbered as $\mathfrak{s}_j, j = 1, \ldots, N$. Following [62], we do not identify nodes of $S$ with individual neurons; instead, these are *neuronal circuits* generating oscillations. The state of each circuit $\mathfrak{s}_j$ is mathematically described by the complex variable $z_j$. Why is it complex? To describe oscillatory dynamics, it is convenient to use two (real) variables $(q, p)$, where $q$ is coordinate (possibly generalized) and $p$ is momentum—the conjugate variable to $q$. By setting $z = (q + ip)/\sqrt{2}$ we move from the real phase space to a complex representation. See Appendix A and article [44] for details.

Oscillations in circuits are random—random oscillations (ROs). For each node-circuit $\mathfrak{s}_j$, ROs are expressed as $\mathbb{C}$-valued random variable $z_j = z_j(\omega)$, where $\omega$ is a random parameter. These random variables are correlated. So, $S$ generates random vector $z = z(\omega) \in \mathbb{C}^N$. The complex linear space $\mathbb{C}^N$ is endowed with the scalar product

$$\langle v | w \rangle = \sum_j^N v_j \overline{w}_j. \qquad (1)$$

This is a complex Hilbert space; it is denoted by the symbol $\mathcal{H}$. Such spaces are basic in quantum theory. So, a complex Hilbert space is naturally coupled to the phase space dynamics [47,44] (see Appendix A).

Geometrically, a neuronal network $S$ can be represented by a graph $G_S$ with nodes given by neuronal circuits $\mathfrak{s}_j, j = 1, \ldots, N$. These nodes are connected by edges. In the simplest case, nodes are individual neurons and edges are axon-dendrite connections between neurons. In this case, the network's graph $G_S$ is a directed multigraph, with the direction of an edge determined by the axon's origin. Signals propagating through the axon-dendrite network generate correlations between ROs in neurons. These correlations play a crucial role in constructing the QL representation of classical neuronal dynamics. Generally, the structure of connections between node-circuits is very complex and not limited to the axon-dendrite network (see the article for detailed analysis); some connections are established at the molecular level or via electromagnetic fields. The construction of the complete connection graph $G_S$ is difficult, if not impossible. Moreover, for real neuronal networks in the brain and body, $G_S$ is a hypergraph and its structure varies with time. (We recall that in a hypergraph, edges connect clusters of nodes rather than individual nodes.) In our modeling, we do not employ this extremely complex geometry of connections within S and instead represent the network state by considering correlations between ROs in node-circuits (but cf. with [53-55] where the graph geometry was explored). Thus, instead of the very complex hypergraph $G_S$ of electrochemical connections between node-circuits, it is useful to represent $S$ by the graph $G_{S,cor}$ of correlations between ROs generated in nodes of $G_S$. This approach leads to the QL representation. The set of its nodes coincides with the nodes of the "electrochemical graph" $G_S$, while its edges are determined by nonzero correlations between nodes: if the correlation between ROs in node-circuits $\mathfrak{s}_j$ and $\mathfrak{s}_i$ is nonzero, these nodes are connected by an edge. Such edges are undirected. Hence, $G_S$ is a directed graph, but $G_{S,cor}$ is an undirected graph.

The canonical basis in the linear space $\mathcal{H}$ consists of the vectors $|1\rangle = (10...0), ..., |N\rangle = (0...1)$. Any vector $v \in \mathcal{H}$ can be expanded with respect to this basis, $v = \sum_j^N v_j |j\rangle$. The basis vectors $(|j\rangle)_{j=1}^N$ represent the node-circuits of the network $S$. The node-circuits of S are represented by orthogonal vectors with respect to the scalar product. This orthogonality is a constraint on the model and, in principle, can be omitted.

This is the place to remark that the representation of network nodes by vectors in complex Hilbert space is a formal mathematical construction—the node-circuits are classical (macroscopic) neuronal structures. The networks under consideration are not quantum networks; they are classical networks for which we construct the QL representation.

Let vector $z = z(\omega) \in \mathcal{H}$ be a random vector representing ROs generated by the neuronal network $S$. We proceed under the assumption that it has a zero mean value, $\mu = E[z] = 0$. If this is not the case, one can always use the vector $(z - \mu)$, which has zero mean value. Consider the covariance matrix of this random vector:

$$C = (c_{km}), \quad c_{km} = E[z_k \bar{z}_m]. \qquad (2)$$

This is the matrix representation of the correlation graph $G_{S,cor}$. It is a Hermitian and positive-definite matrix. Its only difference from a density matrix is that $C$ may have a non-unit trace. It is natural to connect the classical ROs in $S$ with quantum formalism, constructing a QL representation of the functioning of $S$ via trace normalization (see [45-48]),

$$C \to \rho = C / \operatorname{Tr} C. \qquad (3)$$

Such a density matrix is a QL state generated by ROs in $S$.

We speak of matrices to couple QLM to the neuronal basis. We can proceed in the basis-invariant framework with covariance and density operators. Thus, we refer to operators or their matrices. As in quantum theory, various bases in $\mathcal{H}$ can be employed. Bases other than the canonical node basis contain linear combinations of node state vectors, so such states are "non-local" from the perspective of brain geometry, as is the body in three-dimensional physical space. This reflects the nonlocality of information processing.

The correspondence, ROs → covariance matrix, is not one-to-one; a variety of ROs generate the same $C$. Moreover, the correspondence $C \to \rho$ is also not one-to-one because of

normalization scaling. Hence, QLM provides a fuzzy picture of classical random processes in a network.[2]

Now consider the covariance matrix $C_j$ such that only one element $c_{jj} \neq 0$. It expresses ROs in $S$ such that all circuits besides $\mathfrak{s}_j$, are inactive (frozen). (For $i \neq j$, condition $c_{ii} = E[|z_i|^2] = 0$ implies that the random variable $z_i = 0$ almost everywhere). While this is an idealized situation, it remains useful in a mathematical model. The corresponding density matrix represents the projection on the vector $|j\rangle$, $\rho_j = C_j/c_{jj} = |j\rangle\langle j|$. In principle, in this way the circuit-basis vector $|j\rangle$ can be physically generated by activating a single circuit in isolation from others.

Now consider ROs with the covariance matrix $C_v = (c_{ij} = v_i \bar{v}_j)$, where vector $v = (v_1, ..., v_N) \in \mathcal{H}$. Then $C_v = |v\rangle\langle v|$ and $\rho_v = |\psi_v\rangle\langle\psi_v|$, where $\psi_v = v/||v||$. Thus, such ROs generate pure states of this QLM. Due to the degeneration of correspondence, ROs → covariance (density) matrix, each pure state can be generated by a variety of ROs. What is common between such ROs? As was shown in [45-48], each such random vector $z = z(\omega)$ is concentrated in the one-dimensional subspace $L_\psi = \{v = c|\psi_v\rangle\}$. If vector $v$ is a non-trivial superposition of the node-basis vectors ($|i\rangle$), that is $v = \sum_i v_i |i\rangle$, then ROs generating the QL state $\rho_v$ are nonlocally distributed; all neuronal nodes $|i\rangle$ with $v_i \neq 0$ are involved in its generation.

We note that one of the ways to generate a pure state is to consider deterministic (non-random) dynamics in $S$, $z_t$ with $z_0 = |v\rangle$, where $|v\rangle$ is normalized to one (see Appendix A). If this initial vector $|v\rangle$ is the eigenvector of QL Hamiltonian, then $\rho_v(t) \equiv |v\rangle\langle v|$. Thus, stationary pure states, one-dimensional projections, can be generated as eigenstates of Hamiltonian dynamics in $S$ (see Appendix A).

### Ensemble versus time averages

This is a good place to make the following theoretical remark on the mathematical description of correlations. In classical probability model (Kolmogorov 1933, [67]), the elements of covariance matrix (2) are calculated as the integrals

$$c_{km} = \int_\Omega z_k(\omega)\bar{z}_m(\omega)dP(\omega), \qquad (4)$$

where $\Omega$ is a sample space [67] (its points are random parameters or elementary events) and $P$ is the probability measure on $\Omega$.

---

[2] In the real case, a Gaussian distribution is uniquely determined by its covariance matrix and mean value. But in the complex case, only a *circularly symmetric* Gaussian distribution is uniquely determined by its (complex) covariance matrix. The assumption that ROs in neuronal circuits are circularly symmetric Gaussian makes the correspondence ROs → covariance matrix one-to-one. However, there are no bio-physical arguments supporting such an assumption.

However, in experimental research (both in physics and biology) the following time averages are used. For two complex-valued time series $x(t)$ and $y(t)$, the covariance is defined as:

$$\text{Cov}(x, y) = \frac{1}{T}\sum_{t=1}^{T} x(t)\bar{y}(t), \tag{5}$$

where $T$ is the total number of time points (we proceed under the assumption of zero mean values).

The coincidence of ensemble and time averages is a subtle mathematical issue; their equivalence relies on the assumption of ergodicity. This assumption is widely accepted and often applied automatically. In theoretical models, one typically works within the framework of Kolmogorov probability theory, whereas in experimental studies, time averages are generally used. However, the validity of the ergodicity hypothesis in the context of quantum and QL processes remains an open question [68,65]. A detailed discussion of this issue falls beyond the scope of the present paper, and we shall not pursue it further here.

The formula (5) raises the question of the selection of a proper time scale for averaging—that is, the selection of the parameter $T$. In neuroscience, this issue has been discussed, e.g., in [69]-[72].

# 3. Classical vs. quantum realizations of observables on neuronal networks

Let $S$ be a network with $N$ neuronal circuits. ROs in $S$ are represented by a classical random variable $z = z(\omega)$ valued in complex Hilbert space $\mathcal{H}$ of dimension $N$. (Here parameter $\omega$ describes randomness in $S$.)

Consider now a quadratic form $Q_A(z) = \langle Az|z\rangle$ on $\mathcal{H}$, where $A$ is a Hermitian matrix. For a random vector $z$ valued in $\mathcal{H}$, we can consider the average of this form,

$$E_z[Q_A] = \int_\Omega \langle A z(\omega)| z(\omega) \rangle\, dP(\omega) = \int_\mathcal{H} \langle Aw|w\rangle\, dp_z(w), \tag{6}$$

where $\Omega$ is a set of random parameters (elementary events), $P$ is the probability measure, and $p_z$ is the probability distribution of the $\mathcal{H}$-valued random vector $z = z(\omega)$.

This average can be coupled to the covariance matrix by the following equality:

$$E_z[Q_A] = \text{Tr}\, C A. \tag{7}$$

The right-hand side of this equality is (up to normalization) the quantum formula for calculation of averages of observables that are mathematically represented by Hermitian matrices (operators). In terms of the corresponding density matrix

$$\rho = \rho_C = C\,/\,\text{Tr}\, C$$

we have

$$\langle A \rangle_\rho = \text{Tr } \rho A = E_z[Q_A] / \text{Tr } C = \int_{\mathcal{H}} \langle Aw|w \rangle \, dp_z(w) / \int_{\mathcal{H}} \langle w|w \rangle \, dp_z(w). \tag{8}$$

This formula couples the average $\langle A \rangle_\rho$ of quantum observable $A$ in the state $\rho$ with the average of the corresponding quadratic form of ROs in the neuronal network $S$.

The correspondence rule

$$A \leftrightarrow Q_A \tag{9}$$

generates matching of quantum and classical averages. One can investigate this coupling in more detail and obtain from the quadratic form $Q_A = Q_A(\omega)$ a discrete random variable with values $a_i$, where $(a_i)$ is the spectrum of the operator $A$. Such a discrete random variable is obtained via the threshold detection scheme; the Born rule appears as an approximation of classical probability. This is a mathematically advanced formalism that cannot be presented here (see [73] for rigorous mathematical representation). In this model of classical-quantum coupling, the discretization procedure (threshold detection for quadratic forms of classical neuronal variables) is the source of "quantumness." The phase space representation (Section 22 and Appendix A) is purely classical. All quadratic forms are jointly defined as random variables on the same probability space. However, each threshold detection measurement procedure determines its own probability space. Generally, the corresponding discrete-valued observables cannot be represented as random variables on the same probability space. They may not be jointly measurable, as their measurement procedures need not be compatible. In this model, the appearance of incompatible observables originates from the transition from classical observables, given by quadratic forms, to QL observables mathematically represented by Hermitian operators.

Consider now a dichotomous observable $A$ yielding values 0,1. It has two representations, classical and QL. In the QL representation $A = E$ is given by a projector on subspace $L$; in the classical representation $A$ is given by the quadratic form $Q_E$. Let ROs in $S$ are described by the random vector $z$ with the covariance matrix $C$, the QL state $\rho = C / \text{Tr } C$. Since $A$'s average is equal to the probability of the outcome $A = 1$, we have the following coupling between this probability and classical average of $Q_E$,

$$P(A=1|\rho) = E_z[Q_E] / \text{Tr } C = \int_{\mathcal{H}} \langle Ew|w \rangle \, dp_z(w) / \int_{\mathcal{H}} \langle w|w \rangle \, dp_z(w). \tag{10}$$

It can be interpreted as the "weight" of ROs in the subspace $L = E\mathcal{H}$ relatively to the weight of ROs the whole space $\mathcal{H}$. Thus, this formula connects the outcomes of observations over a neuronal network $S$ with averaging of ROs in it.

Generally if $A = \sum_i a_i E_{a_i}$, where $(E_{a_i})$ is the spectral family of the operator $A$, then we have

$$P(A=a_i|\rho) = E_z[Q_{E_{a_i}}] / \text{Tr } C. \tag{11}$$

If operator $A$ has non-degenerate spectrum with eigenvectors $(|a_i\rangle)$, then

$$P(A = a_i|\rho) = c_{ii}/\sum_j c_{jj}, \qquad (12)$$

where $(c_{jj})$ are diagonal elements of the covariance matrix $C$ in the basis $(|a_i\rangle)$. Hence, the probabilities of outcomes are straightforwardly coupled with the elements of the covariance matrix.

Let $A_1$ and $A_2$ be two compatible observables that is they can be jointly measurable. In the QL representation they are described as commuting Hermitian operators $A_1$ and $A_2$. Their quantum correlation is given by the formula

$$\langle A_1 A_2 \rangle_\rho = \text{Tr } \rho\, A_1 A_2 = \text{Tr } \rho\, A_2 A_1. \qquad (13)$$

so, in fact, this is the average of the observable $A$ described by the (Hermitian) operator $A = A_1 A_2 = A_2 A_1$. By applying correspondence rule (9) to this observable, we obtain the classical average representation of quantum correlations

$$\text{Tr } \rho\, A_1 A_2 = E_z[Q_{A_1 A_2}]/\text{Tr } C = \int_\mathcal{H} \langle A_1 A_2 w|w\rangle \, dp_z(w) / \int_\mathcal{H} \langle w|w\rangle \, dp_z(w). \qquad (14)$$

This possibility of a classical representation of quantum correlations may appear to contradict the violation of Bell inequalities. However, it has been shown that this is not the case [47]. Bell-type inequalities can, in principle, be tested for neuronal networks in the brain (as well as for other types of networks, such as social networks). We examined this for observables determined by EEG measurements in article [65].

On classical phase space, one can consider not only quadratic forms but also arbitrary functions, $z \to f(z)$. Can one identify their QL images? In [45-48], the following coupling between classical (functional) and QL (operator) descriptions was considered: $f \to \frac{f''(0)}{2}$ for a twice differentiable function $f = f(z)$.

# 4. Entanglement of observables

In this Section, we present the observational viewpoint on entanglement (see [57-60]). It will be employed in our QLM in Section 5. The traditional approach, viewing entanglement as the correlation of the states of two systems [56], in our case, two neuronal networks $S_1$ and $S_2$, will be employed in Section 6.

We begin with a simple example that illustrates the general construction to be developed later.

Let dim $\mathcal{H}$ =4, let commuting Hermitian operators $A_1$ and $A_2$ have eigenvalues $a_1 = \pm 1, a_2 = \pm 1$, each with degeneracy 2. Consider the basis $(|ij\rangle), i, j = \pm$, in $\mathcal{H}$ consisting of the joint eigenvectors, $A_1|ij\rangle = i|ij\rangle, A_2|ij\rangle = j|ij\rangle$. Any vector in $\mathcal{H}$ can be expanded w.r.t. this basis,

$$|\psi\rangle = \sum_{ij} w_{ij} |ij\rangle. \qquad (15)$$

Such vector decomposition generates on $\mathcal{H}$ the structure of tensor product $\mathcal{H}_1 \otimes \mathcal{H}_2$ where $\mathcal{H}_1$ and $\mathcal{H}_1$ are two-dimensional Hilbert spaces with the bases $(|i\rangle_1, i = \pm)$, and $(|i\rangle_2, i = \pm)$; vector

$$|\phi\rangle = \sum_{ij} w_{ij} |i\rangle_1 \otimes |j\rangle_2 \tag{16}$$

is identified with vector $|\psi\rangle$ given by (15). Tensor product on $\mathcal{H}_1 \otimes \mathcal{H}_2$ induces tensor-product structure on $\mathcal{H}$.

Consider now complex Hilbert space *dim $\mathcal{H}$* $=N =N_1 N_2$. Let commuting Hermitian operators $A_1$ and $A_2$ have eigenvalues $(a_{1j}), j = 1, \ldots, N_1, (a_{2i}), i = 1, \ldots, N_2$. Each $a_{1j}$ has degeneracy $N_2$ and each $a_{2j}$ has degeneracy $N_1$. Consider the basis $(|ij\rangle \equiv |a_{1i}a_{2j}\rangle), i1, \ldots, N_1, j = 1, \ldots, N_2$ in $\mathcal{H}$ consisting of their joint eigenvectors, $A_1|ij\rangle = a_{1i}|ij\rangle, A_2|ij\rangle = a_{2j}|ij\rangle$. Any vector in $\mathcal{H}$ can be expanded w.r.t. this basis, see ([any]). Such vector decomposition generates on $\mathcal{H}$ the structure of tensor product $\mathcal{H}_1 \otimes \mathcal{H}_2$, where $\mathcal{H}_1$ and $\mathcal{H}_2$ are Hilbert spaces of the dimensions $N_1$ and $N_2$ with the bases $(|i\rangle_1)$ and $(|j\rangle_2)$. And isomorphism map $T: \mathcal{H}_1 \otimes \mathcal{H}_2 \to \mathcal{H}$ is determined by its action on the basis vectors $|i\rangle_1 \otimes |j\rangle_2 \to |ij\rangle$.

If the coefficients in representation (15) can be factorized, $w_{ij} = w_i^{(1)} w_j^{(2)}$, then formally vector $|\psi\rangle$ belonging to $\mathcal{H}$ can be written as

$$|\psi\rangle = \left(\sum_i w_i^{(1)} |i\rangle_1\right) \otimes \left(\sum_j w_j^{(2)} |j\rangle_2\right). \tag{17}$$

Such a vector is called separable; otherwise, $|\psi\rangle$ is called entangled. We remark that if $|\psi\rangle = T(|\phi_1\rangle \otimes |\phi_2\rangle)$, then it is factorizable. Thus, the notions of separability versus entanglement given in (17) are equivalent to the usual notions of separability versus entanglement in the tensor product of Hilbert spaces.

Consider now spaces of density operators $D(\mathcal{H}_1), D(\mathcal{H}_2), D(\mathcal{H})$, then $D(\mathcal{H})$ is isomorphic to tensor product $D(\mathcal{H}_1) \otimes D(\mathcal{H}_2)$. The notions of separability vs. entanglement for density operators (mixed states) is transferred to the space $D(\mathcal{H})$.

Denote by the symbol *L(M)* the space of linear operators acting in a complex Hilbert space *M*. We recall that we consider the finite-dimensional case.

In Hilbert space $\mathcal{H}_1 \otimes \mathcal{H}_2$ consider two operator algebras:

$$\boldsymbol{A_1} = \{ A_1 = a_1 \otimes I: a_1 \in L(\mathcal{H}_1) \}, \boldsymbol{A_2} = \{ A_2 = I \otimes a_2: a_2 \in L(\mathcal{H}_2) \}. \tag{18}$$

Hermitian operators belonging to these algebras are called "local observables." For our neurophysiological applications, it is important to note that this is tensor-product locality; generally, it has nothing to do with space-time locality. The images of these algebras in $L(\mathcal{H})$ are denoted as $\boldsymbol{A_i}(\mathcal{H})$, *i=1,2*, or simply $\boldsymbol{A_i}$. These local algebras induce the structure of the tensor product of operators in $\mathcal{H}$; for $A_i \in \boldsymbol{A}_i(\mathcal{H})$, *i=1,2*, we set $A_1 \otimes A_2 = A_1 \circ A_2 = A_2 \circ A_1$. We remark that if $A_1 \in \boldsymbol{A}_1(\mathcal{H})$, $A_2 \in \boldsymbol{A}_2(\mathcal{H})$, then $[A_1, A_2] = 0$; so Hermitian operators from algebras $\boldsymbol{A_1}(\mathcal{H})$ and $\boldsymbol{A_2}(\mathcal{H})$ represent compatible observables.

To clarify the meaning of entanglement (which up to now has been treated from a mathematical viewpoint), consider the four-dimensional case and the singlet state

$$|\psi\rangle = (|+\rangle|-\rangle - |-\rangle|+\rangle)/\sqrt{2}. \tag{19}$$

It is entangled according to the above mathematical definition. But what does it mean physically and biologically (Sections 5, 6)? As noted, this formula encodes correlations between the outcomes of two observables $A_1$ and $A_2$: the conditional probabilities $P(A_2 = \pm|A_1 = \mp) = 1$ as well as $P(A_1 = \pm|A_2 = \mp) = 1$. These correlations, for each pair of such observables, are purely classical. "Quantumness" appears because of the existence of *incompatible observables*, $A_i$, $B_i \in \boldsymbol{A}_i(\mathcal{H})$ such that $[A_i, B_i] \neq 0, i = 1,2$. Here, noncommutativity expresses the impossibility of joint measurements of two observables. The singlet state can also be represented as

$$|\psi\rangle = (|C = +\rangle|D = -\rangle - |C = -\rangle|D = +\rangle)/\sqrt{2}, \tag{20}$$

where $C = A_1, D = A_2$ or $C = B_1, D = B_2$ and the operators $B_i$ can be selected as non-commuting with $A_i, i = 1,2$. Thus, this entangled state encodes correlations between families of local observables that are jointly non-measurable. Classical probability describes only correlations for families of jointly measurable observables. This represents the incompatibility (noncommutativity) interpretation of entanglement.

## 5. Entanglement of observables on neuronal circuits

Here we employ the observational viewpoint on entanglement presented in Section 4. Let $S$ be a network with $N = N_1 N_2$ neuronal circuits. Let $A_1$ and $A_2$ be observables on $S$ as in Section 4. The only new component is that these QL observables are coupled to the network as the QL images of the corresponding quadratic forms $Q_{A_1}$ and $Q_{A_2}$ of ROs in $S$. Neuronal node-circuits $\mathfrak{s}_i, i = 1, \ldots, N$, are renumerated as $\mathfrak{s}_{ij}, i = 1, \ldots, N_1, j = 1, \ldots, N_2$. The biological counterpart of this mathematical construction is that, for each node-circuit, both observables $A_1$ and $A_2$ can be jointly measured, as well as any two observables from the operator algebras $\boldsymbol{A_1}$ and $\boldsymbol{A_2}$.

If a circuit $\mathfrak{s}$ is not activated, then in the classical representation $z_\mathfrak{s} = 0$ a.e., where $z_\mathfrak{s}$ is the random variable describing ROs in $\mathfrak{s}$.

Consider one node-circuit $\mathfrak{s} = \mathfrak{s}_{ij}$. Let only this circuit be activated in the network $S$. In the classical representation, all random variables $z_\mathfrak{s} = 0$ a.e. for $\mathfrak{s} \neq \mathfrak{s}_k$ and $E[|z_k|^2] \neq 0$, where we set $z_k \equiv z_{\mathfrak{s}_k}$. In the QL representation, $S$ is in the pure state $|k\rangle = |ij\rangle$. A measurement of the observables $A_1$ and $A_2$ on the network $S$ gives the outputs $A_1 = a_{1i}$ and $A_2 = a_{2i}$ with probability 1.

Let only two node-circuits be activated in $S$: $\mathfrak{s}_k = \mathfrak{s}_{i_1 j_1}, \mathfrak{s}_m = \mathfrak{s}_{i_2 j_2}$, that is, in the classical representation, $z_n = 0$ a.e. for $n \neq k, m$ and $E[|z_k|^2], E[|z_m|^2] \neq 0$. Let ROs in $\mathfrak{s}_k, \mathfrak{s}_m$ be correlated, generating the covariance matrix $C$ with the elements $c_{kk} = 1, c_{mm} = 1, ´c_{km} = $

$-1, c_{mk} = -1$ and other elements in $C$ equal to zero. Hence, there is perfect anti-correlation between ROs in circuits $\mathfrak{s}_k$ and $\mathfrak{s}_m$. The corresponding QL state $\rho = C/2 = |\psi\rangle\langle\psi|$, where $|\psi\rangle$ is the singlet state (19).

Let $i_1 = +, j_1 = -, i_2 = -, j_2 = +$. The circuits $\mathfrak{s}_{+-}, \mathfrak{s}_{-+}$ generate ROs

$$z = z_{+-}|+-\rangle - z_{-+}|-+\rangle,$$

where $z_{+-} = z_{-+}$ a.e. Thus, the singlet state (19) is the QL image of this random variable. Such correlation is purely classical and does not represent the essence of the QL framework. As was already noted in 4, the strength of employing the QL linear space representation lies in the possibility (for the brain) of using a variety of bases. We have considered the neuronal basis, but the same singlet state carries anti-correlations in various bases (see (20)), whose elements are not localized in specific neuronal circuits.

Formally (mathematically), the neuronal network S can be represented as a compound system $S = (S_1, S_2)$ of two systems $S_1$ and $S_2$ with the state spaces $\mathcal{H}_1$ and $\mathcal{H}_2$. In this observational framework, these systems are not identified with specific neuronal networks (cf. Section 6). They are formally extracted from S with the aid of two algebras of commuting observables, $A_1$ and $A_2$. Measurements performed by observables belonging to $A_i$ are formally treated as "local observables" for subsystem $S_i, i = 1,2$.[3]

The correlations of local observables can be represented as the QL-average. For $B_1 = b_1 \otimes I$ and $B_2 = I \otimes b_2$,

$$\langle b_1 b_2 \rangle_\rho = \text{Tr } \rho \, b_1 \otimes b_2, \rho = \frac{C}{\text{Tr } C}, \quad (21)$$

where $C$ is the covariance operator of ROs in $S$ which are represented by a random variable $z$ valued in tensor product Hilbert space $\mathcal{H} = \mathcal{H}_1 \otimes \mathcal{H}_2$ and not in Cartesian product Hilbert space $\mathcal{H}_1 \oplus \mathcal{H}_2$. As was pointed out in Section 3, such a correlation can be presented in the classical probabilistic framework as

$$\langle b_1 b_2 \rangle_\rho = \frac{1}{\text{Tr } C} = \int_{\mathcal{H}_1 \otimes \mathcal{H}_2} \langle b_1 \otimes b_2 \, w | w \rangle \, d p_z(w). \quad (22)$$

Entanglement is the Hilbert space expression of correlations between local observables—from algebras $A_1(\mathcal{H})$ and $A_2(\mathcal{H})$. The crucial difference from the classical probabilistic representation is that *these algebras contain incompatible observables which cannot be jointly measurable.*

Finally, we stress once again that decomposition of a neuronal network $S$ into subsystems, $S = (S_1, S_2)$, is not physical. Subsystems are virtual, and they express biological functions of $S$, not its neuronal architecture. Decomposition is not unique; even dimensions of the components of the tensor product can vary for the same biophysical neuronal network $S$, say if $N = 12$, we can factorize it as $N_1 = 3, N_2 = 4$ or $N_1 = 2, N_2 = 6$.

---

[3] Although subsystem $S_i$ cannot be identified with a physical network of node-circuits, for external observer (who cannot "open the brain" and see the individual neuronal structure of $S$), subsystems $S_i, i = 1,2$, "exist" and their existence is determined via local observations.

## 6. Entanglement of neuronal states

We start with a recollection of the notion of entanglement for pure and mixed states.

A pure state $|\psi\rangle$ belonging to tensor product of two Hilbert spaces $\mathcal{H} = \mathcal{H}_1 \otimes \mathcal{H}_2$ is called separable if it can be factorized as

$$|\psi\rangle = |\psi\rangle_1 \otimes |\psi\rangle_2, \text{ where } |\psi\rangle_i \in \mathcal{H}_i, \qquad (24)$$

otherwise a pure state is called entangled. A mixed state given by a density operator $\rho$ is called separable if it can be represented as a mixture

$$\rho = \sum_k p_k \, \rho_{k1} \otimes \rho_{k2}, \qquad (25)$$

where $\rho_{ki} \in \mathcal{H}_i$ and the weights $(p_k)$ form a probability distribution, $p_k > 0, \sum_k p_k = 1$. A non-separable state is called entangled. For pure states, it is straightforward to decide whether a state is separable or not. For mixed states, it is very difficult.

Although these definitions are commonly accepted in quantum theory, a careful reading of the original works of Schrödinger [74] may give the impression that he discussed "entanglement of observable" (cf. with discussion in [60]).

Consider now two networks $S_1$ and $S_2$ consisting of neuronal circuits $\mathfrak{s}_{1,j}, j = 1, \ldots, N_1$, and $\mathfrak{s}_{2,j}, j = 1, \ldots, N_2$, respectively. As before, for simplicity, suppose that each circuit generates one complex dimension. So, in QLM for $S_1$ and $S_2$, the complex Hilbert spaces $\mathcal{H}_1$ and $\mathcal{H}_2$ have dimensions $N_1$ and $N_2$. ROs in them are mathematically represented as random vectors $z_i \in \mathcal{H}_i$, i=1,2; here $z_i = (z_{i,1}, \ldots, z_{i,N_i})$.

Consider now the network $S_\oplus$ consisting of node-circuits of $S_1$ and $S_2$ and its graph picturing. The set of nodes of the graph $G_{S_\oplus}$ is the *union* of the sets of nodes of the graphs $G_{S_1}$ and $G_{S_2}$; the set of its edges includes all edges of $G_{S_1}$ and $G_{S_2}$ as well as additional edges representing the connections between some nodes of $G_{S_1}$ and $G_{S_2}$. According to our approach, the edge structure of the graph $G_{S_\oplus}$ is is not visible in the QL representation, which is instead based on the correlation graph $G_{S\_1 \oplus S\_2; \text{cor}}$. The edges of this graph correspond to nonzero correlations between ROs in the corresponding nodes.

In QLM for such a network, its complex Hilbert space $\mathcal{H}_{S_\oplus} = \mathcal{H}_1 \oplus \mathcal{H}_2$ has the dimension $(N_1 + N_2)$. ROs in $S_\oplus$ are mathematically represented the random vector **z**= $(z_1, z_2)$ $\in \mathcal{H}_{S_\oplus}$, so in the node-basis **z**= $(z_1 \ldots z_{N_1+N_2})$. The covariance matrix of ROs has the dimension $(N_1 + N_2)^2$. This dimension does not match the dimension of a density matrix for a quantum compound system, that is $N^2, N = N_1 N_2$. Such a network cannot be used for generating entanglement.

We suggest the following construction of a compound network $S_\otimes$ whose complex Hilbert space is not a direct sum but a tensor product, $\mathcal{H}_{S_\oplus} = \mathcal{H}_1 \oplus \mathcal{H}_2$, and the covariance matrix of ROs in $S_\otimes$ has the dimension $N^2$. Creation of the network $S_\otimes$ that is able to generate entangled states is characterized by the emergence of new circuits not present (or activated) in $S_\oplus$.

Each pair of circuits $\mathfrak{s}_{1,i} \in S_1$ and $\mathfrak{s}_{2,j} \in S_2$ is combined into the new circuit $\mathfrak{s}_{ij}$. How can such a compound circuit be created? Since $\mathfrak{s}_{ij}$ consists of the same neurons as the circuits $\mathfrak{s}_{1,i}, \mathfrak{s}_{2,j}$ the only new structure in the circuit $\mathfrak{s}_{ij}$ arises from generating (activation) new channels for communication between neurons in $\mathfrak{s}_{1,i}$ and neurons in $\mathfrak{s}_{2;j}$. These channels can be physical axon-dendrite connections activated in the network $S_\otimes$ (but not active before). Another testable hypothesis is that *entangling channels* are routes for electromagnetic signaling between neurons across the circuits $\mathfrak{s}_{1,i}$ and $\mathfrak{s}_{2,j}$ [75,66,76,77]. Chemical signaling may also contribute to the formation of the entanglement-generating network $S_\otimes$, albeit on slower time scales.

One can hypothesize that the brain can generate entangled networks using various communication channels to perform different tasks. Generally, the three sorts of communication channels mentioned can be involved in the creation of circuits $\mathfrak{s}_{ij} \in S_\otimes$; each such circuit is given by the triple

$$\mathfrak{s}_{ij} = (\mathfrak{s}_{1,i}, e_{ij}, \mathfrak{s}_{2,j}), \tag{25}$$

where $e_{ij}$ denotes the signaling channel between the neuronal circuits $\mathfrak{s}_{1,i}$ and $\mathfrak{s}_{2,j}$. ROs in this circuit are described by a random variable $Z_{ij} = Z_{ij}(\omega)$, where $\omega$ is a chance parameter. Such ROs generate the covariance matrix $C_{ij;km} = E[Z_{ij}\bar{Z}_{km}]$, whose elements are the correlations between circuits $\mathfrak{s}_{ij}$ and $\mathfrak{s}_{km}$. This matrix has the dimension $(N_1 N_2)^2$. In QLM, the corresponding density matrices are obtained via normalization by trace; in the operator representation, they act in the complex Hilbert space $\mathcal{H}$ of the dimension $N = N_1 N_2$.

We remark that these compound circuits need not be connected at the physical level, e.g., by an axon-dendrite connection. Signals propagating in channels $e_{ij}$ and $e_{km}$ generates electromagnetic fields, and they can be non-trivially correlated (see Section 7 on *ephaptic* generation of such correlations).

Thus, circuits $\mathfrak{s}_{ij}$ are vertices of the graph $G_{S\_1 \otimes S\_2; \text{cor}}$ ; its edges $E_{ij,km}$ connect these vertices and represent correlations between ROs in circuits $\mathfrak{s}_{ij}$ and $\mathfrak{s}_{km}$. We discuss only the graph $G_{S\_1 \otimes S\_2; \text{cor}}$ which edges $E_{ij,km}$ represent correlations between corresponding circuits $\mathfrak{s}_{ij}$ and $\mathfrak{s}_{km}$. The graph of physical connections is not essential for QLM.

What is about the tensor-product structure of the Hilbert space $\mathcal{H}$? Let only one circuit $\mathfrak{s}_{ij}$ is activated and let it generate ROs $Z_{ij}$. In the corresponding (self-)covariance matrix $C_{ij}$ only one element is nonzero, namely, $c_{ij;ij} = E[|Z_{ij}|^2] \neq 0$. In QL, such a covariance matrix is represented by the pure state $|ij\rangle \in \mathcal{H}$ (or the density operator $\rho_{ij} = |ij\rangle\langle ij|$).

Now consider the compound neural network $S = (S_1, S_2)$ with an arbitrary pattern of ROs. In QLM, its covariance matrix $C$ is given by the density matrix

$$\rho = C / \text{Tr } C = \sum_{ij,km} r_{ij,km} |ij\rangle \langle km|.$$

Hence, the state space of density operators acting in $\mathcal{H}$ is represented as the tensor product $D(\mathcal{H})=D(\mathcal{H}_1) \otimes D(\mathcal{H}_2)$, where $\mathcal{H}_i$ is the Hilbert space for the network $S_i, i = 1,2$.

Up to now, we have followed the standard quantum framework for a compound system $S = (S_1, S_2)$. As usual, we can consider the marginal states of $\rho$ generated by partial tracing,

$$\rho_1 = Tr_{\mathcal{H}_2} \rho, \quad \rho_2 = Tr_{\mathcal{H}_1} \rho. \tag{26}$$

In the quantum formalism, these states are interpreted as the states of the subsystems $S_1$ and $S_2$ of the system $S = (S_1, S_2)$. However, in our neuronal model, we can consider the states $\rho_{S_1}$ and $\rho_{S_2}$ of the neuronal networks $S_1$ and $S_2$ in the absence of signaling between them. We remark that the network $S_1 \otimes S_2$ is created via activation of the cross-connection between neurons in $S_1$ and $S_2$ and the inter-network connections contribute to signaling between $S_1$ and $S_2$ only indirectly. In short, the correlations $c_{S_m;ij}/= E[z_{m,i}\bar{z}_{m,j}], m = 1,2$, cannot be reconstructed from the covariance matrix $C$ of correlations in $S_1 \otimes S_2$ and the subsystems' QL states $\rho_{S_m}, m = 1,2$, are not equal to the marginal states $\rho_m$.

We consider the observables $A_1$ and $A_2$. If only the circuit $\mathfrak{s}_{ij}$ is activated, then $A_1 = i, A_2 = j$ with probability 1. In QLM, they are represented by the operators, which are diagonal in the basis $(|ij\rangle)$. Then we can use the construction from Sections 4, 5—observational entanglement. In particular, we create the tensor-product structure,

$\mathcal{H}=\mathcal{H}_1 \otimes \mathcal{H}_2.$

## 7. Ephaptic entanglement

This is an appropriate place to point out *ephaptic coupling between neuronal structures in the brain [75,66,76]*. This coupling enables communication between neurons that differs from direct systems based on physical connections, such as electrical synapses and chemical synapses. Through this mechanism, signals in nerve fibers can be correlated as a result of local electric fields. Ephaptic coupling can generate synchronization of action potential firing in neurons [77].

Recently, this coupling was highlighted in article [78] on modeling nonlocal representation of memories:

*"It is increasingly clear that memories are distributed across multiple brain areas. Such 'engram complexes' are important features of memory formation and consolidation. Here, we test the hypothesis that engram complexes are formed in part by bioelectric fields that sculpt and guide neural activity and tie together the areas that participate in engram*

*complexes. ... Our results ... provide evidence for in vivo ephaptic coupling in memory representations."*

Our QL formalism describes such nonlocal correlations and, in particular, memories distributed across multiple brain areas. Such nonlocal memories are encoded in QL states. Here we again recall our basic conjecture that the brain explores the QL representation, i.e., it operates not with oscillations in neuronal networks but with the corresponding covariance matrices.

Thus, the creation of the entanglement-generating network $S_\otimes$ is *a process*. At each instance in time, the character of signaling between neurons in the circuits $\mathfrak{s}_{1,i}$ and $\mathfrak{s}_{2,j}$ plays a crucial role in the generation of a new circuit $\mathfrak{s}_{ij}$ and a specific state of $S_\otimes$, entangled or not.

# 8. Toward experimental verification of mental entanglement

## 8.1. Experimental framework for entanglement of neuronal networks

Here we consider the simplest case of the model for entanglement of two neuronal networks presented in Section 6 (see also [55]). In this way, we can generate only a restricted set of entangled states (in contrast to the general model). Nevertheless, some examples of entangled states can still be obtained.

The nodes of the compound network $S_1 \otimes S_2$ are given by the pairs of nodes of the networks $S_1$ and $S_2$,

$$\mathfrak{s}_{ij} = (\mathfrak{s}_{1,i}, \mathfrak{s}_{2,j}), \tag{27}$$

and ROs in these node-circuits are described as the random variables $Z_{ij} = z_{1,i} z_{2,j}$, where the random variables $z_{1,i}$ and $z_{2,j}$ describe ROs in node-circuits of corresponding neuronal networks. Hence, the elements of the cross-covariance matrix $C$ are given as

$$c_{ij,km} = E[Z_{ij}\bar{Z}_{km}] = E[z_{1,i}z_{2,j}\bar{z}_{1,k}\bar{z}_{2,m}] \tag{28}$$

Consider now two spatially separated areas in the brain and two sets of electrodes $\mathfrak{s}_{1,i}, i = 1, \ldots, N_1$, and $\mathfrak{s}_{2,j}, j = 1, \ldots, N_2$, coupled to the respective areas. Their outputs correspond to random variables $z_{1,i}$ and $z_{2,j}$. Under the assumption of ergodicity, we can identify statistical averages with time averages. We center the random variables by subtracting their averages. We calculate the cross-covariance matrix. Then, by using a test of entanglement (Section 9), we determine whether the corresponding density matrix represents an entangled or separable state.

Since directly measuring correlations between signals generated by individual neurons in the brain is experimentally complicated (but cf. Section 8.4), it is natural to employ

EEG/MEG techniques to measure correlations between signals generated in spatially and functionally separated brain areas (see Section 8.3 for further discussion of this proposal).

We also mention fMRI as a possible modality, but its limited temporal resolution makes it less suitable for detecting fast or transient correlations as required for entanglement. EEG or MEG would be more appropriate owing to their millisecond-scale resolution.

*Note for implementation.*

Practical EEG/MEG implementation details (preprocessing, spectral estimation, window length $T$, and statistical controls) are summarized in Section 8.2; see also standard references [72,69,71].

## 8.2. EEG/MEG implementation: minimum requirements

Prefer *source-space* analyses and state whether leakage correction was applied (e.g., symmetric orthogonalization [79]). Note that sensor-space signals are linear mixtures of underlying sources; therefore, source-space analyses with leakage correction are preferred wherever possible. Report the reference scheme, artifact handling (e.g., Independent Component Analysis (ICA) for EOG/EMG), and filtering (including notch)[72]. For zero-lag confounds, accompany coherence or Phase-Locking Value (PLV) with imaginary coherence and/or wPLI (with limitations noted [80,81]). These analyses quantify statistical dependence and do not by themselves establish directionality or causation. Specify spectral estimation (Welch or multitaper) and effective degrees of freedom; choose $T$ to cover 5–10 cycles ($\Delta f \approx 1/T$) [69,70,71]. Use matched surrogates (trial shuffling or phase randomization) and correct for multiple comparisons [72]. List reproducibility items: SNR, number of tapers/segments, leakage correction, inverse model or regularization, and whether analyses were performed in sensor or source space.

## 8.3. Parallels between classical EEG-based neurophysiological approaches and QLM of cognition

It is important to emphasize that our QL modeling framework shares methodological parallels with well-established neurophysiological tools used in EEG/MEG-based studies [82,83,84]. Both approaches rely on the analysis of covariance structures and correlations among signals.

### Functional connectivity in neuroscience

Functional connectivity refers to the statistical dependencies between spatially distinct neuronal populations. It is formally defined as the *temporal correlation* between neurophysiological signals recorded at different sites in the brain, such as EEG channels.

Mathematically, common measures of functional connectivity include:

*Covariance.*

For two time series $x(t)$ and $y(t)$, the covariance is defined as:

$$\text{Cov}(x,y) = \frac{1}{T}\sum_{t=1}^{T}(x(t)-\mu_x)(y(t)-\mu_y), \tag{29}$$

where $\mu_x, \mu_y$ are the mean values of $x(t)$ and $y(t)$, and $T$ is the total number of time points.

*Pearson correlation coefficient.*

$$r_{xy} = \frac{\text{Cov}(x,y)}{\sigma_x \sigma_y}$$

where $\sigma_x, \sigma_y$ are the standard deviations of the signals.

*Coherence.*

Coherence measures frequency-specific linear correlations between signals:

$$C_{xy}(f) = \frac{|S_{xy}(f)|^2}{S_{xx}(f)S_{yy}(f)}$$

where $S_{xy}(f)$ is the cross-spectral density, and $S_{xx}(f)$ and $S_{yy}(f)$ are the auto-spectral densities. (*Estimator notes:* Welch or multitaper estimators are commonly used; see [69,70,71,85]).

*Phase-Locking Value( PLV).*

PLV quantifies phase synchronization between two signals:

$$\text{PLV} = \left|\frac{1}{T}\sum_{t=1}^{T} e^{i(\phi_x(t)-\phi_y(t))}\right|$$

where $\phi_x(t)$ and $\phi_y(t)$ are the instantaneous phases, typically extracted using the Hilbert transform or wavelet methods (following [83]).

*Scope of functional connectivity (FC) metrics.*

These metrics quantify statistical dependence but not directionality or causation; directional inferences require separate analyses and explicit assumptions.

*Zero-lag confounds and robust metrics.*

To mitigate volume conduction and common reference effects, the imaginary part of coherency and/or the weighted phase-lag index (wPLI) should be reported alongside classical metrics [80,81]. However, these approaches do not eliminate all leakage in sensor space; whenever possible, analyses should be performed in source space with leakage correction (see Section 8.2).

*Applications.*

FC networks are generated by computing these measures pairwise across brain regions, producing a connectivity matrix that is subsequently analyzed for modularity, hub architecture, and network dynamics. These methods are extensively applied to investigate cognition, neurological disorders, and stimulus-driven responses [86,82,87,72].

For tutorials and discussions of interpretational pitfalls, see.

## Parallels with QLM

In the QL framework, similar mathematical constructs arise naturally. The density matrix or generalized covariance matrix represents probabilistic dependencies among cognitive observables, analogous to FC matrices in EEG studies. Furthermore, off-diagonal terms in QL states encode interference-like effects, comparable to EEG phase synchrony measures such as PLV and coherence.

Thus, the QL formalism extends conventional measures by providing a richer probabilistic interpretation grounded in generalized state representations.

Finally, the concept of entanglement in QL modeling can be compared—at a conceptual level—with strong inter-regional correlations identified in FC analyses, where clusters of brain regions operate as integrated modules. This analogy is heuristic and does not indicate equivalence of constructs.

This is an appropriate place to note that classical signal analysis of brain function frequently employs the analytic signal representation via the Hilbert transform (see, e.g., [85,72,83]). The established use of such complex-valued representations suggests another avenue for QL-style modeling of brain dynamics. We plan to develop such a model in a future paper.

## Summary of parallels

**Key takeaway.** The correspondence between EEG/MEG FC and QL constructs is *conceptual*: both capture dependencies through second-order structure (covariances/coherences vs. density matrix off-diagonals). This analogy is heuristic and does *not* imply equivalence of constructs, measurement units, or causal mechanisms.

*Comparison of EEG/MEG-based methods and QL modeling of cognition (conceptual mapping; not a one-to-one equivalence).*

| Concept | EEG/MEG Neurophysiology | QL Modeling of Cognition |
|---|---|---|
| Covariance | Covariance / Correlation | Density matrix (covariances) |
| Synchrony | Coherence / Phase-locking (PLV) | Interference (off-diagonals) |
| Network Correlations | Functional networks | Entanglement |

**Practical implications.**

- Treat FC metrics as measures of *statistical dependence*, not directionality; use source space with leakage correction and report imaginary coherency/wPLI when applicable (Section 8.2).

- When robust FC modules persist under stringent controls, QL analyses can quantify nonseparability via mixed-state entanglement measures (Section 9); Bell-type tests face signaling constraints (see Appendix B).

### 8.4. In vitro neuronal networks

In vitro neuronal networks are cultures of neurons that replicate key aspects of network structure and function. In such preparations, signals from individual neurons or defined circuits can be recorded directly; connectivity can be patterned, and currents across connections between spatially separated subnetworks can be measured. Although experimentally demanding, these paradigms are feasible and align closely with our framework (cf. Section 8.1).

The experimental testing of QL entanglement in vitro is increasingly practical owing to advances in multi-electrode arrays (MEAs), which allow simultaneous stimulation and recording from dozens to hundreds of sites.

One promising approach is *patterned electrical stimulation* to impose structured correlations between distinct subpopulations—for example, time-locked or phase-modulated sequences delivered to spatially separated regions to create controlled coupling patterns.

Additionally, *pharmacological modulation* offers a complementary route, e.g.:

- **Bicuculline** ($GABA_A$ antagonist) to increase network excitability and enhance synchrony;

- **Carbachol** or other acetylcholine agonists to regulate oscillatory dynamics and increase coherence.

These manipulations can serve to test QL nonseparability by:

1. engineering structured couplings via stimulation or pharmacology,

2. recording the resulting activity with MEAs, and

3. analyzing correlations using QL-inspired criteria (e.g., separability bounds).

Such protocols provide a concrete route toward evaluating QL entanglement while maintaining continuity with established neurophysiological methods. Finally, experimental confirmation of QL nonseparability would support quantum-inspired computation with neuronal networks—QL neuromorphic computing (see [15,54,16,55,44] ).

# 9. Basic quantitative measures of entanglement for mixed states

In quantum information theory, the entanglement of mixed states is quantified using various measures defined through density operators. These measures are critical for characterizing quantum correlations in composite systems described by mixed states. In the following, we summarize mixed-state entanglement measures most relevant for empirical analyses of QL states reconstructed from neural signals. In the context of QLM of cognition, such measures may be applied to examine "mental entanglement" and complex cognitive interdependencies.

## 9.1. Terminology

Throughout this section, 'entanglement' refers to the nonseparability of the QL state $\rho$, constructed from classical neural signals (e.g., EEG/MEG/fMRI-derived time series). We do not suggest microscopic quantum entanglement in brain tissue; outcomes depend on the selected subsystem partition and the measurement basis used to define $\rho$ and the partial transpose.

We start with the definition of the *von Neumann entropy* of a generally mixed state given by a density matrix:

$$S(\rho) = -\mathrm{Tr}(\rho \log \rho)$$

It measures the degree of mixedness of the state; $S(\rho) = 0$ if and only if $\rho$ is a pure state. Hence, it can be used as a measure of the purity of a quantum (or QL) state.

Now let $\rho$ be the density operator of a bipartite system on Hilbert space $\mathcal{H}_A \otimes \mathcal{H}_B$.

## 9.2. Entanglement entropy

For a pure bipartite state $\rho_{AB} = |\psi_{AB}\rangle \langle \psi_{AB}|$, the entanglement entropy is defined as the von Neumann entropy of the reduced state:

$$S_A = S(\rho_A), \quad \rho_A = \mathrm{Tr}_B(\rho_{AB})$$

This quantity measures the degree of entanglement between subsystems $A$ and $B$ for pure bipartite states. For pure global states, entanglement is determined by the entropy of the reduced state: $S(\rho_A) > 0$ if and only if $\rho_{AB}$ is entangled (and $S(\rho_A) = 0$ iff it is a product state). In contrast, $S(\rho) = 0$ or the linear entropy $S_L(\rho) = 1 - \mathrm{Tr}\,\rho^2 = 0$ only confirms that the global state is pure, not whether it is entangled across a given bipartition.

In cognitive and neuronal data, however, pure QL states are not expected: noise, nonstationarity, and averaging typically yield mixed states. Therefore, mixed-state entanglement measures are required.

## 9.3. Negativity and logarithmic negativity

Negativity quantifies entanglement by examining the partial transpose of the density matrix:

$N(\rho) = (||\rho^{T_B}||_1 - 1)/2$

where $\rho^{T_B}$ is the partial transpose with respect to the subsystem $B$ and $||\rho^{T_B}||_1$ is the trace norm of $\rho^{T_B}$.

Logarithmic negativity is defined as:

$E_N(\rho) = \log_2 ||\rho^{T_B}||_1$

These are standard entanglement monotones for mixed states; the partial transpose is taken with respect to the chosen subsystem in the product basis.

## 9.4. Concurrence (Two-Qubit systems)

For two-qubit mixed states $\rho$, concurrence is defined as:

$$C(\rho) = \max(0, \lambda_1 - \lambda_2 - \lambda_3 - \lambda_4)$$

where $\lambda_i$ are the square roots of the eigenvalues (in decreasing order) of:

$$R = \rho(\sigma_y \otimes \sigma_y)\rho^*(\sigma_y \otimes \sigma_y)$$

with $\rho^*$ denoting complex conjugation and $\sigma_y$ the Pauli y matrix.

These measures of entanglement quantify quantum correlations between systems. In quantum theory, there is also a measure used to capture combined classical–quantum correlations.

For two-qubit systems, concurrence corresponds to the entanglement of formation.

## 9.5. Quantum mutual information

For mixed states, quantum mutual information measures total correlations:

$$I(A:B) = S(\rho_A) + S(\rho_B) - S(\rho_{AB})$$

If $I(A:B) = 0$, the two systems are uncorrelated (neither classical nor QL correlations). If $I(A:B) > 0$, there are correlations, but this measure does not distinguish QL from classical components and is not an entanglement monotone.

However, mutual information is not a measure of entanglement, because

- **Nonzero for Separable States:** Even separable (non-entangled) mixed states can yield nonzero mutual information because of classical correlations.

- *Entanglement Monotonicity:* Valid entanglement measures must not increase under local operations and classical communication. Mutual information fails this criterion because it quantifies all correlations, not exclusively quantum ones.

## 9.6. Bell inequalities

We note that the degree of violation of Bell inequalities can serve as indirect evidence for entangled observables; however, such tests in cognitive or neuronal contexts are challenging to implement (see Appendix B for discussion).

# 10. Conceptual differences and added value of the QL framework

While there are important methodological parallels between our QL framework and classical neurophysiological approaches, it is essential to emphasize the fundamental conceptual innovations that distinguish the QL model.

Most classical models in neuroscience—such as *Principal Component Analysis* (PCA), *ICA*, and *Integrated Information Theory* (IIT)—are based on analyzing statistical dependencies or decomposing neural signals into independent or maximally informative components. These approaches often assume linearity, Gaussianity, or specific information-theoretic structures, and they generally function within the framework of classical probability theory.

In contrast, the QL framework introduces the formalism of *operator-based observables* and tensor-product *structures*, adapted from quantum theory but applied to macroscopic neuronal information processing. These mathematical tools allow the formal representation of several fundamental cognitive phenomena, including:

- **Contextuality:** The outcome of cognitive measurements depends on the context, similar to the contextuality of quantum measurements.

- **Incompatibility:** Certain cognitive observables cannot be simultaneously measured or precisely assigned, reflecting uncertainty and complementarity in cognition.

- **Entanglement:** Complex dependencies and holistic cognitive states can be modeled through entangled QL states.

Mathematically, these phenomena are naturally expressed using non-commuting operators and composite Hilbert spaces:

$$\mathcal{H}_{\text{total}} = \mathcal{H}_1 \otimes \mathcal{H}_2$$

where the subsystems correspond to distinct cognitive or neural components.

The density operator (or generalized state) in the QL model encodes both classical correlations and quantum-like correlations (entanglement), extending beyond covariance-based approaches such as PCA and ICA.

*Added value*

By enabling such generalized probabilistic structures, the QL approach formally extends classical theories. It provides novel ways to model:

- Non-classical interference effects in cognition.
- Strong contextual dependencies in decision-making.
- Holistic, system-level processes inaccessible to classical decompositions.

This conceptual generalization bridges neurophysiological signal processing with higher-level cognitive modeling, offering an integrated framework for studying cognition beyond traditional statistical tools.

## 11. Concluding remarks

This paper represents a step toward constructing a conceptual and mathematical bridge between oscillatory cognition—defined as the rhythmic activity of neuronal networks—and QL models of cognition, which have been successfully applied to explain contextuality, interference, and entanglement-like effects in human behavior and decision-making. This bridge relies on a fundamental correspondence: QL mental states are represented by density operators and QL observables by quadratic forms, both of which are mathematically grounded in the covariance structure of ROs in neural circuits. These constructions are developed within the framework of PCSFT, which extends beyond the standard quantum formalism [45,46,47,48].

Previous work [44] has suggested that PCSFT provides a viable interpretational and computational foundation for QL representations of cognitive phenomena. Here, we focus on one of the most conceptually challenging issues: *the formalization of mental entanglement—a cognitive analog of quantum entanglement*—which we consider crucial for modeling integrative mental processes involving distributed and interacting brain networks.

In quantum information theory, entanglement is central to the non-classical correlations underlying quantum computational advantage. While the physical meaning of entanglement remains debated—particularly regarding separability and locality—there has been increasing interest in an alternative formulation: *observational entanglement*. This approach avoids problematic metaphysical assumptions and emphasizes statistical correlations observed through measurements. We adopt this perspective as a more transparent entry point into modeling cognitive entanglement (see Section 5).

We then proceed to explore state entanglement in the QL framework, treating entangled QL states as representations of the joint activity of spatially and functionally separated neuronal assemblies. In this context, *mental entanglement* provides a natural mechanism for feature binding—the brain's capacity to integrate disparate perceptual or cognitive contents (e.g., color, shape, and motion) into unified conscious experiences. This formulation suggests a candidate QL solution to the binding problem, long regarded as one of the central unsolved questions in neuroscience and cognitive science.

Moreover, the introduction of mental entanglement offers a speculative yet potentially fruitful path toward addressing aspects of the *hard problem of consciousness*—specifically, how and why certain brain processes give rise to subjective experience. While our approach does not resolve the hard problem, it aligns with perspectives proposing that consciousness may involve non-classical, globally coherent states that resist decomposition into strictly local mechanisms. If mental entanglement reflects a form of nonlocal coherence in brain function, it may point to a formal structure compatible with integrated information and global workspace theories, enriched by QL formalism.

In cognitive neuroscience, numerous studies have shown that neuronal oscillations—particularly in the gamma, theta, and beta bands—are associated with integrative functions such as memory binding, attentional selection, and conscious access. Our model establishes a bridge between these empirical findings and QL cognitive models by interpreting such oscillatory patterns as classical fields whose covariance structures define the QL states and observables. This offers a testable link between neurophysiological processes and the abstract mathematical structures of QL cognition.

We further hypothesize that entangled QL states derived from ROs may underlie enhanced computational capacities in specific neural subsystems, particularly within the cerebral cortex and hippocampus. This aligns with evidence of high-performance integrative processing in these regions and indicates a deeper role for QL representations in modeling cognitive efficiency.

In Section 8.3, we outline preliminary experimental designs aimed at indirectly detecting signatures of mental entanglement using EEG/MEG methodologies, focusing on correlation structures in neural signals that diverge from classical expectations. Although speculative, these tests are intended to guide future empirical investigations.

In conclusion, the development of the mental entanglement formalism reinforces the broader conjecture that the brain employs QL representations in cognitive processing. This framework opens the door to a deeper interdisciplinary synthesis—integrating neurophysiological data, quantum-inspired mathematical tools, and enduring philosophical questions in consciousness research.

## Acknowledgments

The authors were supported by JST, CREST Grant Number JPMJCR23P4, Japan; A.K. was partially supported by the EU-grant CA21169 (DYNALIFE); M.Y. was partially supported by JSPS KAKENHI Grant (23H04830, 22K18265, 23K22379), JST Moonshot R&D Grant (JPMJMS2295), and MEXT Quantum Leap Flagship Program (MEXT QLEAP) Grant (JPMXS0120330644).

## Appendix A. Coupling symplectic Hamiltonian dynamics and the Schrödinger equation

PCSFT provides a classical foundation for quantum mechanics 45-48]. A central idea is that the Schrödinger equation, which lies at the heart of quantum theory, can be derived from or coupled with *symplectic Hamiltonian dynamics* on a classical phase space. This framework examines the quantum-classical interface from both mathematical and conceptual perspectives. At its core, PCSFT interprets quantum states as labels for statistical ensembles of classical oscillators.

For $N$ classical oscillators, the phase space $\Phi = Q \times P = \mathbb{R}^N \times \mathbb{R}^N$. This corresponds to a real Hilbert space with the scalar product $(\phi_1|\phi_2) = (q_1|q_2) + (p_1|p_2)$, where $\phi_i = \{q_i, p_i\} \in \Phi$. This phase space underpins quantum mechanics with a finite-dimensional state space, the complex Hilbert space $\mathcal{H} = \mathbb{C}^N$, endowed with the scalar product described in Section 2.

Quantum mechanics on physical space $\mathbb{R}^3$ is based on the infinite-dimensional Hilbert space of square-integrable complex-valued functions, essentially the Hilbert space $\mathcal{H} = L^2(\mathbb{R}^3, \mathbb{C})$. The underlying classical phase space is given by $\Phi = L^2(\mathbb{R}^3, \mathbb{R}) \times L^2(\mathbb{R}^3, \mathbb{R})$. The real and imaginary parts of a wavefunction $|\psi\rangle \in \mathcal{H}$ correspond to coordinates and momenta in an infinite-dimensional phase space $\Phi$.

Any phase space can be endowed with a *symplectic structure*. In this setting, a symplectic form $\omega$ is introduced as:

$$\omega(\phi_1|\phi_2) = (\phi_1|J\phi_2),$$

where $J$ is the symplectic structure operator defined as

$$J = \begin{vmatrix} 0 & I_p \\ -I_q & 0 \end{vmatrix}, \tag{30}$$

where $I_q, I_p$ are the unit operators in $Q, P$. In the complex Hilbert space $\mathcal{H}$ corresponding to the phase space $\Phi$ the operator $J$ is represented as the operator of multiplication by $i$ (we remark that $\mathcal{H}$ is complexification of $\Phi$, namely, $\mathcal{H} = Q \oplus iP$.

The Hamiltonian functional $H(\phi)$ generates dynamics via

$$\dot{\phi}(t) = J \nabla H(\phi(t)).$$

When the Hamiltonian functional is *quadratic*, i.e., given by a symplectically invariant quadratic form $H(\phi) = (\phi|H\phi)$, the evolution reduces to the linear Schrödinger equation:

$$i \frac{d\psi}{dt}(t) = H\psi(t)$$

Thus, the Schrdïnger equation appears not as a postulate, but as a dynamical law for classical phase space dynamics under a symplectic structure.

In PCSFT, a quantum state (wavefunction or density operator) corresponds to the covariance operator of classical oscillations. Quantum averages emerge from classical statistical averages over this ensemble, allowing an interpretation in which quantum randomness is epistemic—arising from incomplete knowledge of underlying classical oscillations. In this way, quantum mechanics can be understood as a *projection or statistical encoding* of a deeper classical theory, with the Schrödinger equation derived from the Hamiltonian flow of an underlying symplectic system.

We now briefly present the above consideration with $q, p$ coordinates. Any Hamiltonian function $H(q, p)$ generates the system of Hamiltonian equations

$$\dot{q} = \frac{\partial H}{\partial p}(q, p), \ \dot{p} = -\frac{\partial H}{\partial q}(q, p). \tag{31}$$

Consider now a quadratic and symplectically invariant Hamiltonian function

$$H(q, p) = 1/2[(Rp, p) + 2(Tp, q) + (Rq, q)], \tag{32}$$

where $R$ is a symmetric operator, $R^\star = R$ and $T^\star = -T$. The operator (the Hessian of the Hamilton functions)

$$H = \begin{vmatrix} R & T \\ -T & R \end{vmatrix}, \tag{33}$$

Which commutes with the symplectic operator $J$. This is a system of harmonic oscillators, and it can be rewritten as the Schrödinger equation.

# Appendix B. Experimental framework for entanglement of observables in neuronal circuits and Bell test

The observational approach to entanglement is comparatively simpler, as it does not require reconstructing the QL state of a compound neuronal network through the full calculation of its cross-correlation matrix.

Instead, we aim to identify a pair of jointly measurable observables $A_1$ and $A_2$, associated with neuronal circuits forming parts of a network $S$. The most widely studied method for detecting entanglement is the Bell test, which evaluates violations of Bell-type inequalities. Among these, the CHSH inequality is the most frequently applied and provides a natural framework for detecting and quantifying entanglement, where the degree of entanglement is reflected in the magnitude of CHSH inequality violation.

*Interpretational note.* A violation of a Bell inequality rules out a class of local realistic models under specific assumptions (e.g., no-signaling and measurement independence), but it does not, by itself, establish directionality or causation among neural processes, nor does it exclude classical common-drive confounds.

To apply the CHSH test, one must define two pairs of observables, $A_1, B_1$ and $A_2, B_2$. Within each pair, the observables must be incompatible, meaning they are not jointly measurable. Across pairs, the observables must be compatible, meaning they can be jointly measured. Correlations between cross-pair observables are then computed to test for CHSH inequality violations.

However, the structure of observables in neuronal circuits—especially in terms of their compatibility and incompatibility—remains largely unexplored. In QLM, the emergence and nature of incompatible observables is still under active investigation. Current approaches often reduce this question to testing for the order effect [18,19], where the sequence of measurements influences the outcome. Detection of such an effect is typically interpreted as evidence of incompatibility between QL observables.

Yet, this interpretation relies on the assumption that observables are of the projection (von Neumann) type. Within the broader framework of quantum instruments, it is possible to observe the order effect even when the corresponding observables are jointly measurable [92]. This complicates the direct use of the order effect as an indicator of incompatibility in general settings.

At present, there is no clear methodology for identifying suitable candidate observables for the CHSH test in neuronal systems.

Moreover, even in physics, Bell-type experiments posed significant theoretical and technical challenges. The basic Bell framework [91] is affected by various so-called loopholes, and it was only in 2015 that loophole-free experiments were successfully performed. These landmark experiments ultimately led to the awarding of the 2022 Nobel Prize in Physics to Alain Aspect, John Clauser, and Anton Zeilinger.

In cognitive psychology and decision-making research, Bell-type inequalities have also been explored, beginning with early foundational and experimental studies [24] (see for later experiments [6, 25-29]). As in physics, these investigations are complex both conceptually and empirically. In fact, the situation in cognitive domains may be even more challenging because of *the apparent impossibility of eliminating signaling effects*. Signaling—referring to the presence of direct influences between measurement settings and outcomes—complicates the interpretation of experimental data. When present, signaling requires a significantly more sophisticated theoretical framework (see [94] ), which lies beyond the scope of this article.

Moreover, these investigations have been complemented by theoretical developments that bridge probabilistic modeling and conceptual compositionality. Bruza et al. [95], for example, introduced a probabilistic framework to study how meanings of conceptual combinations emerge, showing that quantum-inspired probabilistic models can effectively account for observed non-classicalities in meaning composition. Together, these contributions indicate that Bell-type inequalities and contextuality analyses are not merely metaphors but tools with empirical and explanatory power in cognitive science. However, fully accounting for signaling effects and developing corresponding theoretical models remains an open challenge in the field.

Given these challenges, the experimental framework discussed in Sections 8.1 and 8.3 appears more feasible for near-term implementation.

---

i